\documentstyle[aps,graphics]{revtex}

\def\r{{\bf r}}
\def\k{{\bf k}}

\def\q{{\bf \hat q}}
\def\dq{\frac{d^3{\bf q}}{(2\pi)^3}}
\def\ie{{\it i.e.}}
\def\m{{\bf m}}
\def\ez{{\bf \hat e}_z}
\def\re{{\rm Re}}
\def\im{{\rm Im}}

\begin{document}
\draft
\title{Effective index of refraction, optical
rotation, and circular dichroism in isotropic chiral liquid
crystals}
\author{D. Lacoste$^1$ and P. J. Collings$^2$ and T. C. Lubensky$^1$}
\address{$^1$ Department of Physics,
University of Pennsylvania, Philadelphia, PA 19104, USA \\
$^2$ Department of Physics and Astronomy, Swarthmore College,
Swarthmore, PA 19081, USA}
\date{\today}
\maketitle
\begin{abstract}
This paper concerns optical properties of the isotropic phase
above the isotropic-cholesteric transition and of the blue phase
BP III. We introduce an effective index, which describes spatial
dispersion effects such as optical rotation, circular dichroism,
and the modification of the average index due to the fluctuations.
We derive the wavelength dependance of these spatial dispersion
effects quite generally without relying on an expansion in powers
of the chirality and without assuming that the pitch of the
cholesteric $P$ is much shorter than the wavelength of the light
$\lambda$, an approximation which has been made in previous
studies of this problem. The theoretical predictions are supported
by comparing them with experimental spectra of the optical
activity in the BP III phase.
\end{abstract}

\section{Introduction}
Chirality in liquid crystals produces a fascinating variety of
phases, such as the blue phases. Three blue phases (BPs)
designated BP I, BP II, and BP III have been identified, and their
structures are now well understood. BP I and BP II exhibit
long-range periodic order at the half micron scale and Bragg
scatter visible light. For these reasons, the optical study of
blue phases is an active field of research, with a particular
emphasis on spatial dispersion effects such as the optical
activity.

The first experiments on the optical activity in the
pretransitional region of the isotropic phase above the
isotropic-cholesteric transition were carried out by Cheng and
Meyer \cite{cheng}. Using a general formulation due to de Gennes,
they calculated and confirmed experimentally that
 the pretransitional optical activity depends on the temperature as $(T-T^*)^{-0.5}$,
where $T^*$ is the metastability temperature of the isotropic
phase. Two years later, Brazovskii and Dmitriev developed the
first complete theory of phase transitions in cholesteric liquid
crystals, and they predicted the existence of the blue phases
\cite{brazovskii}. The optical activity in the pretransitional
region was derived using this approach by Dolganov et al.
\cite{dolganov1}. At this time, a detailed Landau theory of the
cholesteric blue phases was obtained by Hornreich and Shtrikman,
who also provided an outstanding study of the light scattering in
the blue phases including a treatment of the polarization of the
light, based on the formalism of the M\"{u}ller matrices
\cite{hornreich1,hornreich2}. Bensimon, Domany, and Shtrikman
studied the optical activity in the pretransitional regime and in
the blue phases \cite{bensimon}, confirming and extending the work
of Dolganov. In their paper, the blue phase BP III was considered
as an amorphous polycrystalline structure distinct from the
isotropic phase, and their treatment of the optical activity
relied on the long wavelength approximation. On the experimental
side, the wavelength dependence of the optical activity in the
blue phases was measured by Collings \cite{collings2,collings3}.
The structure of BP III still remained mysterious until
experiments clearly showed a continuous transition between BP III
and the isotropic phase \cite{collings4,keyes} and the existence
of a critical point terminating a line of coexistence of the two
phases \cite{collings5}. The same year, Lubensky and Stark
developed a theory for the isotropic to BP III transition with a
chiral liquid-gas like critical point \cite{lubensky}, which was
extended to include scaling theory in Ref.~\cite{collings8}. Very
recently, there has been a renewal of interest in the study of the
blue phases with the discovery of the smectic blue phases
\cite{pansu}, exhibiting isotropic-BP III coexistence terminating
at a critical point \cite{jamee}.

In section \ref{sec:theory} of this paper, we introduce an
effective index for the propagation of light in the isotropic
phase and in BP III, which contains pretransitional effects such
as the optical activity and the circular dichroism. In section
\ref{sec:long_wavelength}, we present a standard derivation which
is based on the long wavelength approximation \cite{bensimon}. In
section \ref{sec:de Vries}, we make a digression on the optical
properties of periodic chiral media for comparison with the case
of non-periodic chiral media. By evaluating our weak-scattering
expression for the optical dielectric constant tensor in a
cholesteric phase, we find the well-known de Vries formula, which
describes the optical activity of light propagating in this medium
along the pitch axis. In section \ref{sec:exact_wavel} we come
back to isotropic chiral media and obtain the optical activity for
arbitrary wavelength and chirality. This approach is new in that
it is general, not limited to the long wavelength approximation,
and applicable in particular to the resonant region where the
wavelength of the light is of the order of the pitch of the
cholesteric, a regime which has not been considered in detail in
previous studies. In the long wavelength approximation, we collect
contributions to the optical activity up to order $(P/\lambda)^4$.
Though various contributions up to this order have appeared
separately in the literature
\cite{brazovskii,dolganov1,bensimon,dolganov2} and very recently
in \cite{sing}, these references are not fully consistent with
each other. We therefore think that it is important to collect and
discuss all these contributions in one place. We then generalize
our method and obtain the wavelength dependance of the circular
dichroism and of the symmetric part of the index, which to our
knowledge have not appeared in the literature and have never been
measured. Finally in section \ref{sec: Theory&exp}, our
theoretical predictions are supported by experimental spectra of
the optical activity for different values of the chirality in the
BP III phase \cite{collings7}. The agreement between theory and
experiments is very good. Experiments confirm the presence of a
broad maximum in the magnitude of the optical activity when the
wavelength is of the order of the pitch and for positive
(respectively negative) pitch. This feature is the pretransitional
signature in isotropic chiral media of the divergence arising from
the long-range periodicity of the cholesteric phase, BP I and BP
II.

\section{Theory}
\label{sec:theory} Let $\epsilon$ be the relative complex-valued
dielectric constant of the cholesteric liquid crystal in the
isotropic phase or in the blue phase BP III with respect to the
average dielectric constant:
\begin{equation}
\epsilon_{ij}(\r)=\delta_{ij} + \delta \epsilon_{ij}(\r),
\end{equation}
where $\delta \epsilon_{ij}(\r)$ is a relative anisotropic small
local fluctuation of the dielectric constant: $\langle \delta
\epsilon_{ij}(\r) \rangle=0$, $|\delta \epsilon_{ij}(\r)| \ll 1$,
and $\langle..\rangle$ denotes a thermal average. To simplify the
notations, we take the velocity of light in the medium to be $1$,
and we do not indicate explicitly the dependence of the dielectric
constant on the frequency of light in the medium denoted $\omega$.
Maxwell equations in the absence of sources lead to the following
Helmholtz equation for the electric field ${\bf E}(\r,\omega)$:
\begin{equation}
\label{helmoltz} \nabla \times \left( \nabla \times {\bf
E}(\r,\omega) \right) = \omega^2 \epsilon \cdot {\bf
E}(\r,\omega).
\end{equation}
From Eq.~(\ref{helmoltz}), we deduce that the propagation of light
in a homogeneous medium in the absence of fluctuations is
characterized by the following Green's function
\begin{equation}\label{Green_libre}
G_{ij}^0 (\k,\omega)=\frac{\Delta_{ij}}{ -\omega^2 + k^2 - i0^{+}}
- \frac{k_i k_j}{\omega^2 k^2},
\end{equation}
where $\Delta_{ij}=\delta_{ij}-k_i k_j /k^2$ is a projector on the
space transverse to the wave vector $\k$. The first term of
Eq.~(\ref{Green_libre}) is the transverse part describing
travelling wave solutions of Maxwell equations and the second term
is the longitudinal part which describes non-propagating modes.
The scattering by the randomly fluctuating part of the dielectric
function $\delta \epsilon_{ij}(\r)$ is described by the following
4-rank tensor \cite{stark}
\begin{equation}\label{def_tensor_B}
B_{ijkl}(\r)=\omega^4 \langle \delta \epsilon_{ik}(\r) \delta
\epsilon_{jl}(0) \rangle.
\end{equation}
Unless specified otherwise, we consider in this paper only
non-absorbing media in which the tensor $B_{ijkl}(\r)$ is real.
The averaged Green's function $G(\k,\omega)$ follows from Dyson's
equation,
\begin{equation}\label{Dyson}
G^{-1}=(G^{0})^{-1} -\Sigma.
\end{equation}
In the weak scattering approximation,
 the tensor $\Sigma$ can be
calculated using the following equation \cite{kats}:
\begin{equation}\label{Sigma}
\Sigma_{ij}(\k,\omega)=\int \dq B_{ijkl}({\bf q}) G_{kl}^0({\bf
k-q},\omega),
\end{equation}
in terms of the Fourier transform $B_{ijkl}({\bf q})$ of the
tensor defined in Eq.~(\ref{def_tensor_B}). Note that
Eq.~(\ref{Sigma}) is general and applies also to periodic media
for which $B_{ijkl}({\bf q})$ has delta-function peaks in which
case this equation takes the form of a discrete sum over Bloch
waves \cite{galatola}.

The general form of $\Sigma(\k,\omega)$ in an isotropic chiral
medium is:
\begin{equation}\label{def_sigma12}
\Sigma_{ij}(\k,\omega)=\Sigma_0(k,\omega)\delta_{ij} +
\frac{i\epsilon_{ijl}{k_l}}{\omega} \Sigma_1(k,\omega) +
\Sigma_2(k,\omega) \frac{k_i k_j + k_j k_i}{2\omega^2}.
\end{equation}
If $\kappa$ denotes a chiral parameter of the medium,
$\Sigma_{ij}(\k,\omega,\kappa)$ satisfy the general symmetry
relation (valid in any medium in the absence of a magnetic field
\cite{landau})
\begin{equation}\label{symmetry_general}
\Sigma_{ij}(\k,\omega,\kappa)=\Sigma_{ji}(-\k,\omega,\kappa),
\end{equation}
and the chiral symmetry relation
\begin{equation}\label{chiral_symmetry}
\Sigma_{ij}(\k,\omega,\kappa)=\Sigma_{ij}(-\k,\omega,-\kappa).
\end{equation}
The chiral symmetry relation implies that $\Sigma_0$ and
$\Sigma_2$ are odd functions of $\kappa$ and that $\Sigma_1$ is an
even function of $\kappa$. The dispersion relation for light
propagation modified by fluctuations is obtained from $\det \left[
G_0^{-1}(\k,\omega) -\Sigma(\k,\omega) \right]=0$. In a basis
composed of two vectors perpendicular to $\k$, the diagonal
elements of $\Sigma$ are equal to $\Sigma_0$, and the off-diagonal
elements are $\pm ik \Sigma_1 /\omega$. In such a basis,
longitudinal terms such as the last term in
Eq.~(\ref{def_sigma12}) vanish,
 and the following relation is obtained
\begin{equation}\label{dispersion}
\omega^2-k^2-\Sigma_0(k,\omega)=\pm \frac{k}{\omega} \Sigma_1(k,\omega).
\end{equation}
This relation determines $k^+(\omega)$ and $k^-(\omega)$ as a
function of the frequency $\omega$. Writing $k=\omega+\delta k$
and expanding Eq.~(\ref{dispersion}) to first order in $\delta k$,
one finds
\begin{equation}
\label{disper} \delta k^{\pm} = \frac{\Sigma_0(\omega,\omega) \pm
\Sigma_1(\omega,\omega)} {-2\omega \mp
\Sigma_1(\omega,\omega)/\omega \mp \partial \Sigma_1/\partial k -
\partial \Sigma_0/\partial k}.
\end{equation}
To first order in $\Sigma$, this gives $k^{\pm}(\omega)= \omega -
\left[ \Sigma_0(\omega,\omega) \pm \Sigma_1(\omega,\omega)
\right]/2\omega$. The optical activity $\Phi$ is defined as the
angle through which the polarization vector of linearly polarized
light has turned when traversing a medium of length $L$. It is
given by $\Phi=\re (k^+(\omega)-k^-(\omega))L/2$. Similarly the
circular dichroism, measuring the difference in transmission from
left and right circularly polarized waves, is $\Psi=\im
(k^+(\omega)-k^-(\omega))L/2$. To first order in $\Sigma$, we have
therefore
\begin{equation}\label{Eq:Phi}
\Phi=\re \frac{\Sigma_1(\omega,\omega)L}{2\omega},
\end{equation}
and
\begin{equation}\label{Eq:Psi}
\Psi=\im \frac{\Sigma_1(\omega,\omega)L}{2\omega}.
\end{equation}
Note that these relations are valid at arbitrary frequency
$\omega$ when Eq.~(\ref{Sigma}) is valid. The fluctuations also
modify the symmetric part of $\Sigma$, \ie, the average dielectric
constant. The relative change of the average dielectric constant
within the same approximation is $\Delta \epsilon_0=-
\Sigma_0(\omega,\omega)/\omega^2$.

\begin{figure}
{\par\centering
\resizebox*{3.6in}{3.6in}{\rotatebox{0}{\includegraphics{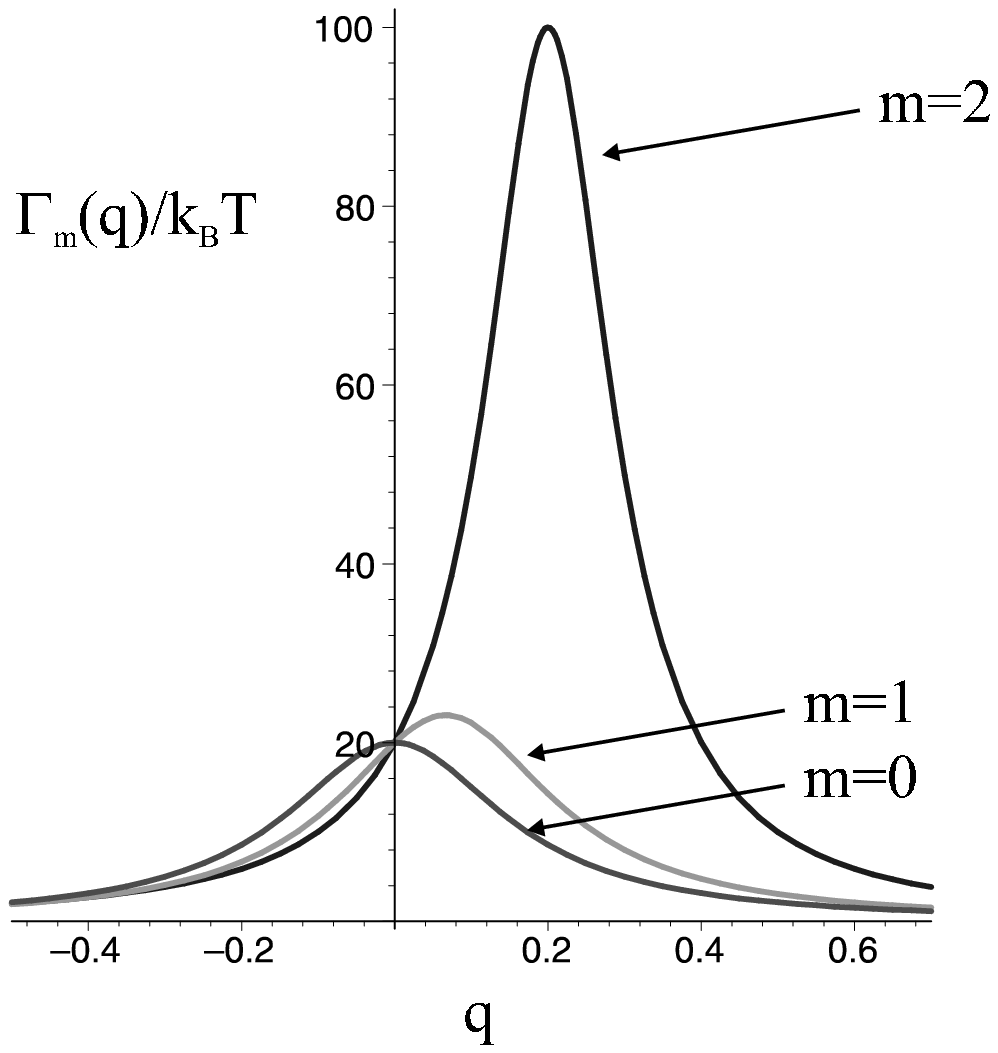}}}
\par}
\caption{Amplitude $\Gamma_m(q)/k_B T$ of the dielectric
anisotropy correlation tensor as a function of the dimensionless
wave vector $q$ for the modes $m$=0,1, and 2. $\kappa$ is $0.2$,
$\rho=1$, and the normalized temperature is $t=0.05$. The curves
show the predominance of the mode $m=2$, because of the choice
$t>t_2>t_1$ and $t$ close to $t_2$.} \label{Fig:Gamma}
\end{figure}

We use in this paper the form of the Landau-de Gennes free energy
of Ref.~\cite{hornreich2},
\begin{equation}\label{landaufreeenergy}
F=V^{-1} \int d^3 \r \{ \frac{1}{2} \left[ a \epsilon_{ij}^2 + c_1
\epsilon_{ij,l}^2 + c_2 \epsilon_{ij,i} \epsilon_{lj,l} -2d
e_{ijl} \epsilon_{in} \epsilon_{jn,l} \right] -\beta \epsilon_{ij}
\epsilon_{jl} \epsilon_{li} + \gamma (\epsilon_{ij}^2)^2 \},
\end{equation}
where $\epsilon_{ij,l}=\partial \epsilon_{ij} /
\partial x_l$ and $e_{ijk}$ is the Levi-Civita fully antisymmetric tensor.
 The coefficient $a$ is proportional
to the reduced temperature $t$, whereas all other coefficients are
assumed to be temperature independent. As explained in
Refs.~\cite{hornreich2} and \cite{lubensky}, it is convenient to
write this free energy in dimensionless form, by expressing all
lengths in units of the order parameter correlation length
$\xi_R=(12\gamma c_1/\beta^2)^{1/2}$. The chirality is then
$\kappa_0=q_c \xi_R$ where $q_c=d/c_1$ is the wave number of the
cholesteric phase which determines the pitch $P=4 \pi /q_c$. The
reduced temperature is defined by $t=(T-T_N^*)/(T_N-T_N^*)$ where
$T_N^*$ is the temperature corresponding to the limit of
metastability of the isotropic phase \cite{tom}, and
$\rho=c_2/c_1$ is the ratio of the two Landau coefficients, which
is of order 1. Using the Gaussian approximation for the
fluctuations \cite{bensimon}, the dielectric anisotropy
correlation tensor defined in Eq.~(\ref{def_tensor_B}) takes the
following form
\begin{equation}\label{tensor_B}
B_{ijkl}({\bf q})=\omega^4 \sum_{m=-2}^{m=2} \Gamma_m(q) T_{ik}^m
(\q) T_{jl}^{-m} (\q),
\end{equation}
where $\Gamma_m(q)$ is evaluated from the equipartition theorem
\begin{equation}\label{def_Gamma}
\Gamma_m(q)=\frac{k_B T}{t-m\kappa q +q^2 \left[ 1+ \frac{\rho}{6}
(4-m^2) \right] }.
\end{equation}
Note that $q$ denotes the dimensionless wavevector measured in
units of $1/\xi_R$, and that Eq.~(\ref{def_Gamma}) has been
corrected from the expression given in Ref.~\cite{lubensky} (the
factor $\rho/4$ has been replaced by $\rho/6$ in agreement with
Ref.~\cite{hornreich2}). In Eq.~(\ref{def_Gamma}), the parameter
$\kappa$ represents an effective chirality at the temperature $t$.
Except at the critical point, $\kappa$ is different from
$\kappa_0$, and the difference $\kappa-\kappa_0$ is proportional
to the order parameter of the BP
 III-isotropic phase transition \cite{lubensky}. The index $m$
 in Eq.~(\ref{tensor_B}) denotes the five independent
modes of the symmetric and traceless part of the dielectric tensor
$\delta \epsilon_{ij}(\r)$. Note that Eq.~(\ref{def_Gamma}) is
equivalent to
\begin{equation}\label{deftm}
\frac{k_B T}{\Gamma_m(q)}=t-m\kappa q + \Delta_m q^2  = \Delta_m
(q-q_m)^2+\tau_m,
\end{equation}
with $\Delta_m=1+ \frac{\rho}{6} (4-m^2)$, $q_m=m
\kappa/2\Delta_m$, $t_m=\Delta_m q_m^2$ and $\tau_m=t-t_m$. We
have introduced $t_m$, the transition temperature of the mode $m$,
and $q_m$, the wavevector which minimizes the energy of the mode
$m$. Denoting $r=1+\rho/2$, we have explicitly
$t_1=t_{-1}=\kappa^2/4r$ and $q_1=-q_{-1}=\kappa /2r$ for the
modes $\pm1$, $t_2=t_{-2}=\kappa^2$ and $q_2=-q_{-2}=\kappa$ for
the modes $\pm2$, $t_0=0$ and $q_0=0$ for the mode $0$. A plot of
$\Gamma_m(q)$ as a function of $q$ is shown in
Fig.~\ref{Fig:Gamma} for the modes $m=0,1$ and $2$, a chirality
$\kappa=0.2$ and a temperature $t=0.05$. This particular plot
shows the dominance of the mode $m=2$, because of the choice of
the temperature $t>t_2>t_1$ and $t$ close to $t_2$. More generally
it can be shown that as $t \rightarrow t_m$, $\Gamma_m(q)$ becomes
a Dirac function localized at $q=q_m$.

The tensor $T_{ij}^m$ introduced in Eq.~(\ref{def_tensor_B}) are
eigenvectors of the tensor $B_{ijkl}({\bf q})$, with eigenvalues
$\omega^4 \Gamma_m(q)$ \cite{brazovskii}. These tensors are
\begin{equation}\label{def_T0}
T^0(\q)= \frac{1}{\sqrt{6}} \left( 3 \q \q -1 \right),
\end{equation}
\begin{equation}\label{def_T1}
T^1(\q)= \frac{1}{\sqrt{2}} \left[ \q \m(\q) + \m(\q) \q
\right]=T^{-1}(\q)^*,
\end{equation}
\begin{equation}\label{def_T2}
T^2(\q)= \m(\q)\m(\q) =T^{-2}(\q)^*,
\end{equation}
with $\m=(1/\sqrt{2}) ({\bf \hat \xi} + i {\bf \hat \eta})$ and
$({\bf \hat \xi},{\bf \hat \eta},\q)$ forming a right-handed
system of orthonormal vectors. We choose the vectors ${\bf \hat
\xi}$ and ${\bf \hat \eta}$ such that ${\bf \hat \xi}(-\q)={\bf
\hat \xi}(\q)$ and ${\bf \hat \eta}(-\q)=-{\bf \hat \eta}(\q)$, so
that
\begin{equation}\label{m}
\m(-\q)=\m(\q)^*.
\end{equation}

\section{Dispersion effects in the long wavelength approximation}
\label{sec:long_wavelength} In this section the optical activity
and the circular dichroism of isotropic chiral media is derived
using the method of Ref.~\cite{bensimon}, which is valid in the
long wavelength approximation and to linear order in $\kappa$. In
section \ref{sec:exact_wavel}, we will present another derivation
of these results, which does not rely on an expansion in
$1/\lambda$ or in $\kappa$. We present here the method of
Ref.~\cite{bensimon} because it has been widely used in the
literature. Following Ref.~\cite{bensimon}, we use an expansion to
first order in ${\bf k}$ of the Green's function
\begin{equation}\label{Green_exp}
G^0_{ij}(\k - {\bf q},\omega) \simeq G^0_{ij}({\bf q},\omega)
-k_l \frac{\partial G^0_{ij}({\bf q},\omega)}{\partial q_l},
\end{equation}
with
\begin{equation}\label{DG}
\frac{\partial G^0_{ij}({\bf q})}{\partial q_l}= -\frac{q_i
\Delta_{jl}+q_j \Delta_{il}}{\omega^2 (\omega^2-q^2+i0^+)}+
 \frac{2q_l \Delta_{ij}}{(\omega^2-q^2+i0^+)^2}.
\end{equation}
Using this expression in Eq.~(\ref{Sigma}), we obtain
$\Sigma_1(k\rightarrow0,\omega)$ which was defined in
Eq.~(\ref{def_sigma12}). After integration over $\q$, we obtain
\begin{equation}\label{SigmaR}
\Sigma_1(k\rightarrow0,\omega)=\int dq q^2 {\frac
{{\omega}^{3}q\left[-\Omega_1(q){\omega}^{2}+\Omega_1(q){q}^{2}-
2\,\Omega_2(q){\omega}^{2}\right]}{ 12{\pi
}^{2}\left({\omega}^2-{q}^2+i0^+ \right)^{2}}},
\end{equation}
where $\Omega_m(q)=\Gamma_m(q)-\Gamma_{-m}(q)$. In the integration
of Eq.~(\ref{SigmaR}), we take the upper limit of integration over
$q$ to be infinity. The continuous model used in this paper is
valid only up to wave vectors $q_{max}=2\pi/a$,
 where $a$ is an intermolecular distance, but
 in the case of Eq.~(\ref{SigmaR}), the
 integral is not sensitive to the value of $q_{max}$
 as discussed in Refs.~\cite{dolganov1} and \cite{lubensky}.

\begin{figure}
{\par\centering
\resizebox*{4.6in}{3in}{\rotatebox{0}{\includegraphics{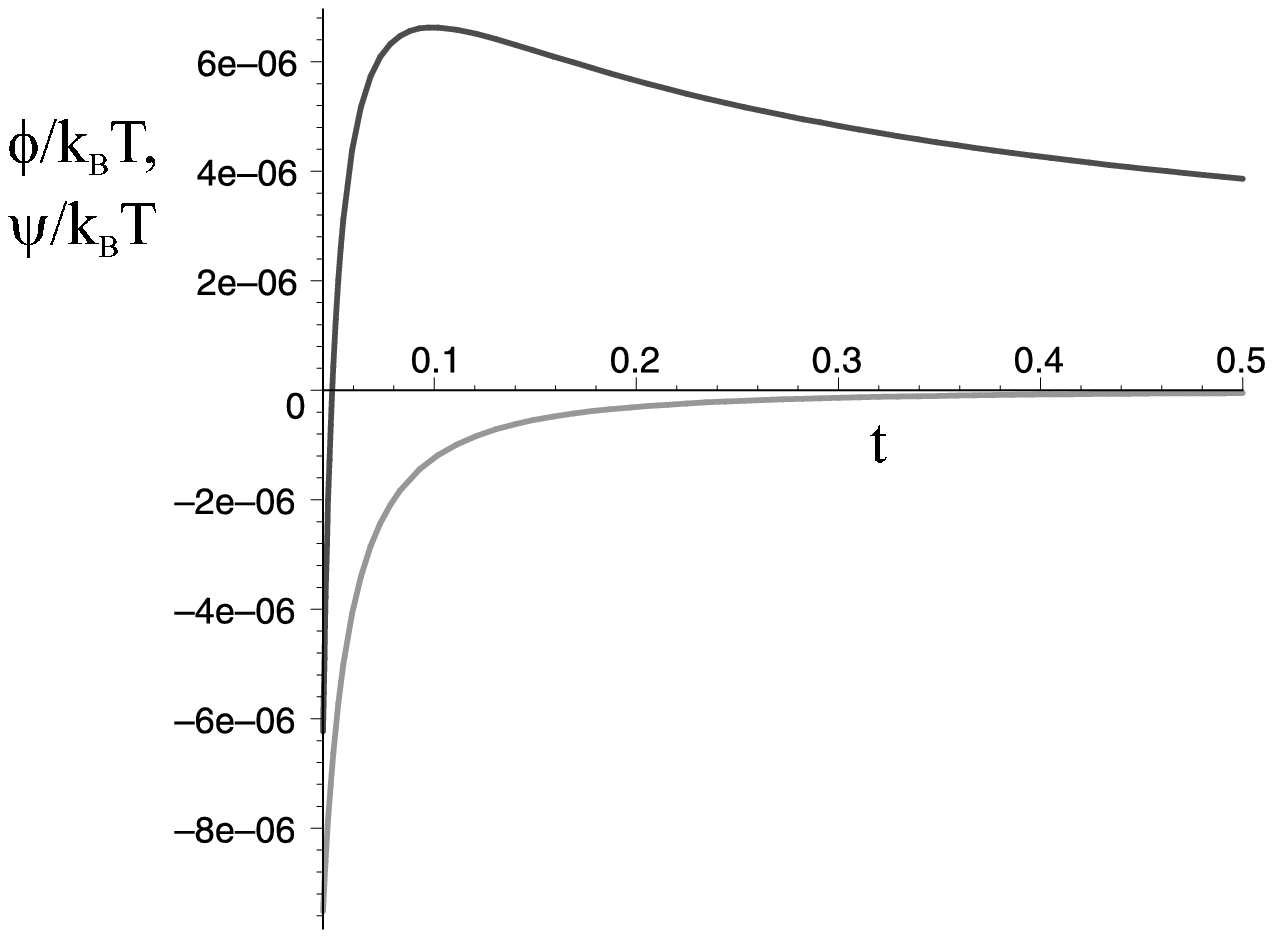}}}
\par}
\caption{Optical activity $\Phi/Lk_BT$, the dark curve, and
circular dichroism $\Psi/Lk_BT$, the light curve as a function of
the temperature $t$ for $\kappa=0.2$. The wavelength is
$\lambda=100  \xi_R$, which corresponds to the long wavelength
regime where Eq.~(\ref{AO_lw}) is applicable. The curve of the
optical activity illustrates the competition between the modes
$m=\pm1$ and $m=\pm2$, the latter been responsible for the
decrease (and the change of sign) of the optical activity when $t$
approaches $t_2$.} \label{Fig:AOvsT}
\end{figure}

Expanding Eq.~(\ref{SigmaR}) in powers of $\omega$ and integrating
over $q$, we obtain the complex-valued $\Sigma_1(\omega)$ in the
long wavelength limit. Using Eqs.~(\ref{Eq:Phi}) and
(\ref{Eq:Psi}), we deduce the optical activity
\begin{equation}\label{AO}
\frac{\Phi}{L}={\frac {\kappa\,\omega^{2} k_B T}{ 48 r^{3/2}\sqrt
\tau_1\pi }}+ \left (\,{\frac {1}{4\sqrt r
\,\tau_1^{3/2}}}-\,{\frac {1}{\tau_2^{3/2} }} \right
)\frac{\omega^{4} k_B T \kappa}{12 \pi},
\end{equation}
and the circular dicroism
\begin{equation}\label{CD}
\frac{\Psi}{L}= \left (-{\frac 1 {\tau_2^{2}}}+\,{\frac 1{6
\,\tau_1^{2}}}\right ) \frac{\omega^{5} k_B T \kappa}{4\pi}.
\end{equation}
Deriving Eqs.~(\ref{AO}) and (\ref{CD}), we have assumed that
$\kappa\ll \sqrt{\tau_2}$ and $\kappa\ll \sqrt{4r\tau_1}$, and the
equations are valid for temperatures such that $t_1<t_2<t$. Within
these approximations, the optical rotation and circular dichroism
are proportional to the chirality $\kappa$. The next order will be
of order $\kappa^3$ as imposed by the symmetry relation of
Eq.~(\ref{chiral_symmetry}). The first term in Eq.~(\ref{AO}) is
identical to the expression of Refs.~\cite{bensimon} and
\cite{dolganov1} for the optical activity in the pretransitional
region. This term is associated with the modes $m=\pm1$ and gives
the $(T-T_1^*)^{-0.5}$ dependence observed in many experiments. As
first noted by Filev, the modes $m=\pm 2$ need to be considered in
addition to the contribution from the modes $m=\pm 1$, in a region
very close to the transition in highly chiral liquid crystals
\cite{filev}. This is the origin of the third term in
Eq.~(\ref{AO}), whose contribution has the opposite sign with
respect to the contribution from the $m=\pm1$ modes. This term
produces a change of sign of the optical activity close to the
transition as shown in Fig.~(\ref{Fig:AOvsT}). There has been some
debate in the literature concerning the temperature dependance of
this term, whether it should be in $(T-T_2^*)^{-1/2}$ or
$(T-T_2^*)^{-3/2}$. The very short paper of Ref.~\cite{filev} did
not provide enough explanations to answer this point, and that of
Ref.~\cite{dolganov2} also lacks explanations and contains some
errors or misprints (in particular in Eq.~(4) of
Ref.~\cite{dolganov2}). In Ref.~\cite{collings1}, Collings et al.
studied experimentally systems of high chirality and found good
agreement with Filev's model with a $1/2$ exponent. Unfortunately
due to a large number of fitting parameters, the data could also
have been explained with an exponent $-3/2$ \cite{collings2}. We
shall come back to this point in section \ref{sec:exact_wavel}. In
a study of the wavelength dependence of the optical activity in
the long wavelength regime \cite{collings3}, Collings et al.
confirmed that the modes $m=\pm 2$ contribute to higher order than
the modes $m=\pm1$ in agreement with Eq.~(\ref{AO}). The presence
of a second order correction in $\omega^4$ for the modes $m=\pm1$
with a dependence $(T-T_1^*)^{-3/2}$, which would correspond to
the second term in Eq.~(\ref{AO}) has not been proved or disproved
experimentally, although quite good fits to the data can be
obtained with this term included as found in Refs.~\cite{sing} and
\cite{sing2}.

\section{De Vries formula}
\label{sec:de Vries} We present in this section a derivation of
the de Vries formula, which is the well known solution of
Maxwell's equations for light propagating in a cholesteric liquid
crystal along the pitch axis \cite{deGennes}. The reason of this
digression into the optics of periodic chiral media will become
clear in the next section, where we explore the relation between
optical activity in isotropic chiral liquid crystals and the de
Vries formula for the cholesteric phase. We shall take the
$z$-axis to be the helical axis of the cholesteric phase, and
(${\bf \hat e}_x$, ${\bf \hat e}_y$, $\ez$) to be a right handed
frame. In the cholesteric phase, the order parameter, the
anisotropy in the dielectric constant, is
\begin{equation}\label{cholesteric_orderparam}
\delta \epsilon_{ij}= \epsilon_a \left( n_i n_j - \frac{1}{3}
\delta_{ij} \right),
\end{equation}
where $\epsilon_a=\epsilon_\parallel-\epsilon_\perp$, and $n=\cos
(\kappa z /2) {\bf \hat e}_x +\sin (\kappa z /2) {\bf \hat e}_y$.
Using Eq.~(\ref{cholesteric_orderparam}) and the definitions
(\ref{def_T0}-\ref{def_T2}), it is simple to show that
\begin{equation}\label{cholesteric_eps}
\delta \epsilon_{ij}(\r)= \delta
\epsilon_{ij}(z)=\frac{\epsilon_a}{2} \left[ e^{i\kappa z}
T_{ij}^{-2} (\ez) + e^{-i\kappa z} T_{ij}^{2} (\ez) \right] -
\frac{\epsilon_a}{\sqrt{6}} T_{ij}^0 (\ez).
\end{equation}
From this, the tensor $B_{ijkl}(\r)=B_{ijkl}(z)$ defined in
Eq.~(\ref{def_tensor_B}) can be constructed. To calculate its
Fourier transform $B_{ijkl}({\bf q})$, it is convenient to use a
spherical coordinate system with ${\bf q}=q(\sin \theta \cos
\phi,\sin \theta \sin \phi, \cos \theta)$,
\begin{equation}\label{cholesteric_B}
B_{ijkl} \left( {\bf q} \right) =\omega^4 \int dx e^{i q \sin
\theta \cos \phi x} \int dy e^{i q \sin \theta \sin \phi y} \int
dz e^{i q \cos \theta z} \delta \epsilon_{ik}(z) \delta
\epsilon_{jl} (0),
\end{equation}
which can be simplified using Eq.~(\ref{cholesteric_eps})
\begin{eqnarray} \label{cholesteric_B_simple} \nonumber
B_{ijkl} \left( {\bf q} \right) & = & \frac{\omega^4 \epsilon_a^2
\left( 2\pi \right)^3}{4 q^2 \sin \theta} \delta(\phi) \delta
(\theta) \left[ \delta \left( q+\kappa \right) T_{ik}^{-2}
T_{jl}^{2} + \delta \left( q-\kappa \right) T_{ik}^{2} T_{jl}^{-2}
\right] + A_{ijkl} ({\bf q}), \\
& =& \frac{\omega^4 \epsilon_a^2 \left( 2\pi \right)^3}{4} \left[
\delta^3 \left( {\bf q}+\kappa \ez \right) T_{ik}^{-2} T_{jl}^{2}
+ \delta^3 \left( {\bf q}-\kappa \ez \right) T_{ik}^{2}
T_{jl}^{-2} \right] + A_{ijkl} ({\bf q}),
\end{eqnarray}
where $T_{ij}^{\pm 2}=T_{ij}^{\pm 2}(\ez)$, and $A_{ijkl}$ is a
linear combination of tensors of the form $T_{ik}^m T_{jl}^{m'}$
with $m$ and $m'$ being $0,2$ or $-2$. The Bragg peaks of the
cholesteric phase located at ${\bf q}=\pm \kappa \ez$ correspond
to the first two terms of Eq.~(\ref{cholesteric_B_simple}). Note
that in a periodic medium considered in this section, $\Sigma_1$
is a function of $\k$, whereas in an isotropic medium to be
considered in the next section, $\Sigma_1$ is only a function of
$k=|\k|$. In a periodic medium $\Sigma_1(\k,\omega)$ can be
derived generally from $\Sigma_{ij}(\k,\omega)$ using
\begin{equation}\label{Delta_Sigma}
\frac{\Sigma_1(\k,\omega)}{\omega}=\frac{ \left[
\Sigma_{ij}(\k,\omega)- \Sigma_{ij}(-\k,\omega) \right]
e_{ijl}k_l}{4ik^2}.
\end{equation}
Using Eqs.~(\ref{Sigma}), (\ref{cholesteric_B_simple}) and
(\ref{Delta_Sigma}), we find
\begin{equation}\label{Sigma1_Vries}
\frac{\Sigma_1(\k,\omega)}{\omega}=\frac{\omega^4 \epsilon_a^2
e_{ijm}k_m}{16 i k^2} \left[ G_{kl}^0 \left( \k + \kappa \ez
\right) - G_{kl}^0 \left( -\k + \kappa \ez \right) \right] \left[
T_{ik}^{-2} T_{jl}^{2} - T_{ik}^{2} T_{jl}^{-2} + \tilde{A}_{ijkl}
\right],
\end{equation}
where $\tilde{A}_{ijkl}= \left( -T_{ik}^{-2} + T_{ik}^{2} \right)
T_{jl}^0 / \sqrt{6}$. In Eq.~(\ref{Sigma1_Vries}), the tensors $T$
can be eliminated using Eqs.~(\ref{def_T0}), (\ref{def_T2}) and
(\ref{Relation1}).

We now assumes that $\k$ is along the $z$-axis. In this case, it
is simple to show that the tensor $\tilde{A}_{ijkl}$ does not
contribute to $\Sigma_1$. According to the analysis of section
\ref{sec:theory}, to linear order in $\Sigma$ the optical activity
and the circular dichroism are proportional to the real and
imaginary part of $\Sigma_1(\k,\omega)$ evaluated at
$|\k|=\omega$. With these assumptions, Eq.~(\ref{Sigma1_Vries})
gives
\begin{equation}\label{Sigma1_Vries3}
\frac{\Sigma_1(\k=\omega \ez,\omega)}{\omega}=\frac{\omega^4
\epsilon_a^2 \kappa}{\left( \kappa^2 + 2 \omega \kappa -i0^+
\right) \left( 2\omega \kappa - \kappa^2 + i0^+ \right)}.
\end{equation}
The real part of Eq.~(\ref{Sigma1_Vries3}) leads to de Vries
formula \cite{deGennes}
\begin{equation}\label{Vries_f}
\frac{\Phi}{L} = \frac{\pi \epsilon_a^2}{16 P \lambda'^2
\left[1-\lambda'^2 \right]},
\end{equation}
where $\lambda'=\lambda/P$ and $P=4 \pi \xi_R/\kappa$, the pitch
of the cholesteric phase, and the circular dichroism $\Psi$ is
zero within the same approximations. Note the following features
\cite{deGennes}: there is a dispersion anomaly at the Bragg
reflection $\lambda'=1$, but both Eq.~(\ref{Vries_f}) and $\Psi=0$
break down near the Bragg reflection. Indeed close to the Bragg
reflection, terms of higher order in $\Sigma$ contribute to the
optical rotation and the circular dichroism, and for this reason
Eqs.~(\ref{Vries_f}) and $\Psi=0$ are only valid in the domain
$\epsilon_a \lambda \ll ||P|-\lambda|$ \cite{kats2}. The sign of
the optical rotation is such that a right-handed helix
($\kappa>0$) produces a positive optical rotation (the material is
laevogyric) when $\lambda \ll P$ and a negative one (the material
is dextrogyric) when $\lambda \gg P$. In the long wavelength limit
$\lambda \ll P$, the de Vries optical activity $\Phi/L$ is of
order $P^3/\lambda^4$ and depends on the light propagation
direction as opposed to the isotropic-BP III phases where it is of
order $P/\lambda^2$ and is independent of the light propagation
direction.

\section{Wavelength dependance of spatial dispersion
effects in an isotropic chiral medium} \label{sec:exact_wavel} In
this section, we compute the exact wavelength dependance of the
optical activity, the circular dichroism and the average index in
an isotropic chiral medium, without relying on the long wavelength
approximation and at any order in the chirality $\kappa$. The
method has been pioneered by Dolganov \cite{dolganov1,dolganov2}
in his study of BP I and BP II, but this reference does not
present all the details of the calculation. Instead of using the
expansion Eq.~(\ref{Green_exp}), which is only valid in the long
wavelength approximation, we calculate exactly
$\Sigma_1(k,\omega)$ from Eqs.~(\ref{Sigma}) and
(\ref{Delta_Sigma}). Separating the contribution from the modes
$m=2$ and $m=1$, we obtain (see appendix \ref{AppendixA} for
details)
\begin{equation}\label{Vries}
\frac{\Sigma_1^{m=\pm2}(k,\omega)}{\omega}=\int \dq \frac{\mp q
\omega^2 c^2
 \Gamma_{\pm2}(q) \left( 2\omega^2-k^2 + k^2 c^2 \right)} {2
\left(k^2+2k q c +q^2-\omega^2-i0^+ \right) \left( k^2-2kqc
+q^2-\omega^2 -i0^+ \right) },
\end{equation}
and
\begin{equation}\label{noVries}
\frac{\Sigma_1^{m=\pm1}(k,\omega)}{\omega}=\int \dq \frac{\mp q
\omega^2 \Gamma_{\pm1}(q) \left( c^2 -1 \right) \left( \omega^2+
4k^2 c^2 - k^2 -q^2 \right)} {4 \left(k^2+2kqc+q^2-\omega^2-i0^+
\right) \left( -k^2+2kqc-q^2+\omega^2+i0^+ \right) },
\end{equation}
with $c={\bf \hat k} \cdot \q$. From these equations, the
contribution of the modes $m=\pm1$ and $m=\pm2$ of
Eq.~(\ref{SigmaR}) of section \ref{sec:long_wavelength} is
recovered in the limit $\k \rightarrow 0$. According to the
analysis of section \ref{sec:theory} however, the correct
procedure to obtain the optical activity and the circular
dichroism to leading order in $\Sigma$ is from
$\Sigma_1(k,\omega)$ evaluated at $k=\omega$. As will become clear
later, this does not give the same result in general when compared
to section \ref{sec:long_wavelength} where the limit $\k
\rightarrow 0$ is taken in Eq.~(\ref{Green_exp}). Using
$k=\omega$, Eqs.~(\ref{Vries}) and (\ref{noVries}) can be
simplified
\begin{equation}\label{Vries2}
\frac{\Sigma_1^{m=\pm2}(\omega,\omega)}{\omega}=\int \dq \frac{\mp
\omega^4 c^2
 \Gamma_{\pm2}(q) \left( 1 + c^2 \right)} {2
\left(q + 2\omega c -i0^+ \right) \left( q-2 \omega c -i0^+
\right)q },
\end{equation}
and
\begin{equation}\label{noVries2}
\frac{\Sigma_1^{m=\pm1}(\omega,\omega)}{\omega}=\int \dq \frac{\mp
\omega^2 \Gamma_{\pm1}(q)}{4q} \left( c^2 -1 \right).
\end{equation}
Note that the integrand of Eq.~(\ref{noVries2}) vanishes for $\k
\parallel {\bf q}$ but is non-zero otherwise. This means that the
modes $m=\pm1$ contribute only if the light is propagating off
axis or if the pitch axis and the director are not perpendicular
to each other as in SmC$^*$ for instance. Eq.~(\ref{Vries2}) is
the analog for isotropic chiral media of the de Vries formula of
Eq.~(\ref{Sigma1_Vries3}) valid for periodic media. The two
expressions become identical with the replacements $c\rightarrow1$
since the light is propagating along the pitch axis, and
$q\rightarrow \pm\kappa$ according to
Eq.~(\ref{cholesteric_B_simple}). Other periodic and chiral media
like BP I and BP II can be treated as the cholesteric phase in
section \ref{sec:de Vries}. Therefore we expect in BP I and BP II
a divergence of the $m=\pm2$ contribution to the optical activity
at the Bragg condition, {\it i.e.} when the denominator of
Eq.~(\ref{Vries2}) is zero, and a change of sign of the optical
activity when crossing this point. There should be no divergence
for the modes $m=\pm1$ as can be inferred from
Eq.~(\ref{noVries2}).

The recent reference of Hunte and Singh \cite{sing} contains a
derivation of the pretransitional optical activity in isotropic
chiral media, which uses a method similar to the one presented in
this article although the authors have only applied it to the long
wavelength regime. We believe that the authors have made an error
in deriving Eqs.~(30) and (31) of Ref.~\cite{sing} which represent
respectively the contribution of the modes $m=\pm1$ and of
$m=\pm2$ to the optical activity. The integrand of Eq.~(30) of
Ref.~\cite{sing} should vanish when $c=1$ as is true for the
integrand of Eq.~(\ref{noVries2}) but it does not, and Eq.~(31) of
Ref.~\cite{sing} should reproduce the de Vries as is true for
Eq.~(\ref{Vries2}) but does not. Therefore we think that the
results of the derivation of the optical activity in
Ref.~\cite{sing} are incorrect although the method used is valid.

A straightforward evaluation of Eq.~(\ref{noVries2}) gives
$\Sigma_1$ for the modes $m=\pm1$ and for an arbitrary frequency
$\omega$:
\begin{equation}\label{Mode1}
\frac{\Sigma_1^{m=\pm1}(\omega,\omega)}{\omega}=\frac{\omega^2
\kappa k_B T}{24 \pi r^{3/2} \sqrt{\tau_1}}.
\end{equation}
The $m=\pm 1$ contribution of the optical activity has a trivial
frequency dependance of $\omega^2$. The reason is the following:
the modes $m=\pm1$ unlike the modes $m=\pm 2$ are longitudinal,
therefore these modes are associated with the longitudinal part of
the Green's function (also called near-field part) which is the
second term in the r.h.s of Eq.~(\ref{Green_libre}). The only pole
of the near-field Green's function is at $\omega=0$, which is the
reason for the absence of a complex wavelength dependance for the
modes $m=\pm1$ and also the reason for which these modes give the
leading contribution to the optical activity at small $\omega$.

Therefore the most interesting part of the wavelength dependance
of the optical activity comes from the modes $m=\pm 2$. It is also
the dominant contribution when $t$ is close to $t_2$. The
integration in Eq.~(\ref{Vries2}) can be done analytically for
$t>t_2$, and the final result (see appendix \ref{AppendixB} for
details of the derivation) takes the following form
\begin{equation}\label{OAG}
\re
\frac{\Sigma_1^{m=\pm2}(\omega,\omega)}{\omega}=\frac{-\omega^3}{32
\pi \sqrt{\tau_2}} \left[ x_1 f(x_1) - x_3 f(x_3) \right] k_B T,
\end{equation}
where $x_1=(\kappa+i\sqrt{\tau_2})/2\omega$ and
$x_3=(-\kappa+i\sqrt{\tau_2})/2\omega$. Note that Eq.~(\ref{OAG})
agrees with the symmetry relation of Eq.~(\ref{chiral_symmetry}).
We have introduced the function
\begin{equation}\label{function_f}
f(x)=-2x^2-\frac{8}{3}+(x+x^3)\ln \left( \frac{x+1}{x-1} \right).
\end{equation}
This function tends to $-8/3$ as $x\rightarrow\infty$, is
equivalent to $16/15x^2$ as $x\rightarrow0$, and diverges at
$x=\pm 1$ as a result of the divergence of the initial expression
of Eq.~(\ref{Vries2}). Eqs.~(\ref{OAG}) and (\ref{function_f})
fully agree with Filev's results in Ref.~\cite{filev} quoted for
$\tau_2=0$. Similar functions but not exactly identical functions
are present in Refs.~\cite{brazovskii}, \cite{dolganov2},
\cite{dolganov3}, and \cite{sing} which could indicate misprints
or errors regarding this particular point.

It is interesting to consider two particular cases of
Eq.~(\ref{OAG}), the case where $\omega \ll \kappa$, which defines
the long wavelength approximation and the opposite case where
$\omega \gg \kappa$. In the first case, it is easy to deduce from
Eq.~(\ref{OAG}) that $\re \Sigma_1^{m=\pm2}(\omega,\omega)/\omega=
-2\omega^4 \kappa k_B T /15\pi \sqrt{\tau_2} t$. Therefore the
optical activity in the long wavelength approximation for a
temperature $t>t_2>t_1$ is
\begin{equation}\label{AO_lw}
\frac{\Phi}{k_B T L}=\frac{\omega^2 \kappa} {48\pi r^{3/2}
\sqrt{\tau_1}} -\frac{\omega^4 \kappa}{15\pi \sqrt{\tau_2}
t}+o(\omega^4).
\end{equation}
Note that the contribution of the modes $m=\pm2$ goes as
$(T-T^*_2)^{-3/2}$ if $\kappa \ll t$, but should go as
$(T-T^*_2)^{-1/2}$ otherwise. As discussed before,
Eq.~(\ref{AO_lw}) is not identical with Eq.~(\ref{AO}) because
$\lim_{\omega \rightarrow 0} \Sigma(\omega,\omega) \neq \lim_{k
\rightarrow 0} \Sigma(k,\omega)$. Since the modes $m=\pm1$ and
$m=\pm2$ contribute with different signs in Eq.~(\ref{AO_lw}),
there is a competition between these two modes, which leads to a
maximum in the optical activity as a function of the temperature
$t$ as shown in Fig.~\ref{Fig:AOvsT} for a chirality $\kappa=0.2$.
In the opposite case of $\omega \gg \kappa$, we find using
Eq.~(\ref{OAG}) that $\re \Sigma_1^{m=\pm2}(\omega,\omega)/\omega=
\omega^2 \kappa k_B T/12\pi \sqrt{\tau_2}$.

\begin{figure}
{\par\centering
\resizebox*{4.6in}{3.2in}{\rotatebox{0}{\includegraphics{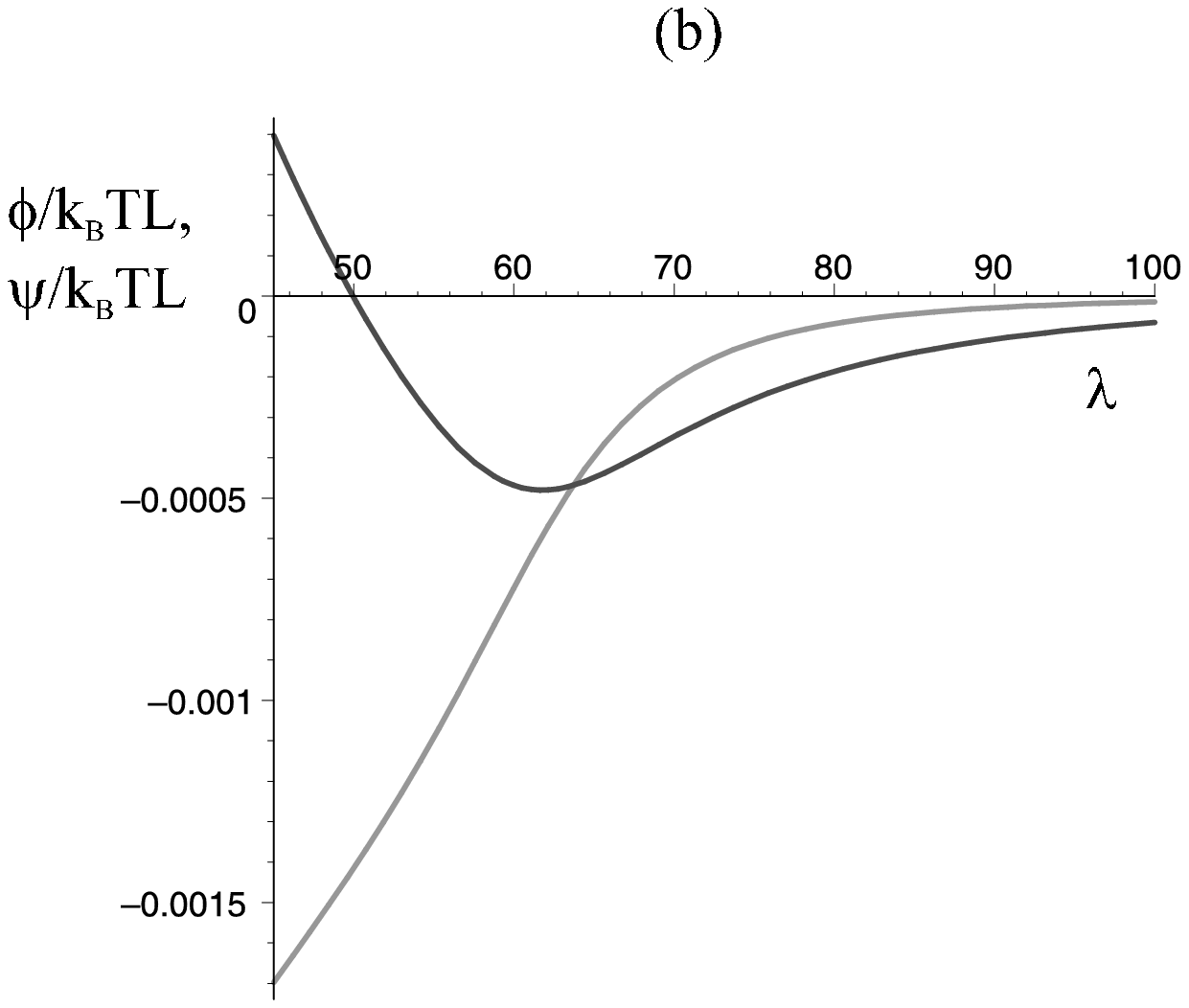}}}
\par} {\par\centering
\resizebox*{4.6in}{3.2in}{\rotatebox{0}{\includegraphics{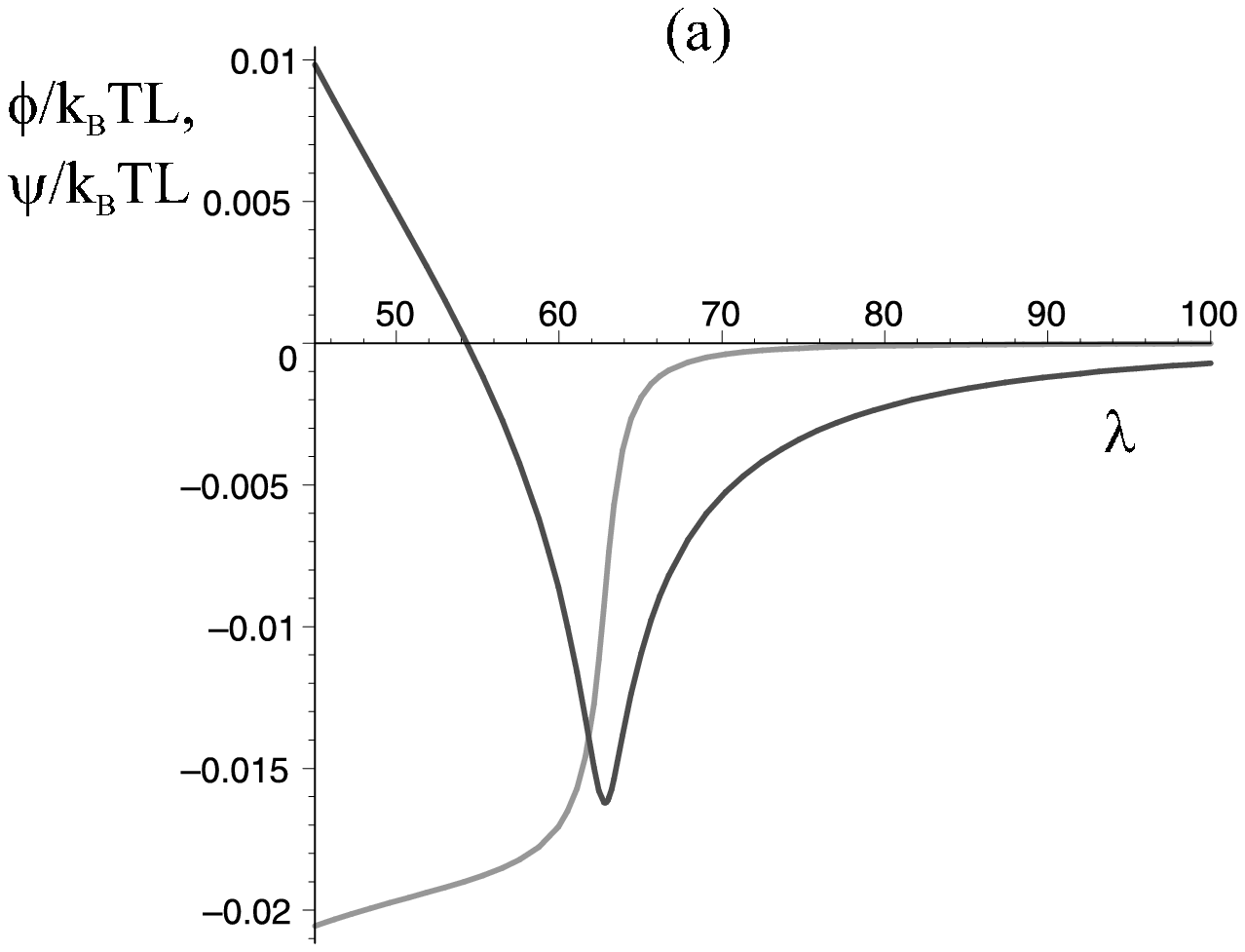}}}
\par} \caption{The contribution from the modes $m=\pm2$ to the
optical activity $\Phi/Lk_BT$, the dark curve, and the circular
dichroism $\Psi/Lk_BT$, the light curve,
  are shown as function of the wavelength $\lambda$ for $\kappa=0.2$.
(a) The temperature $t$ is such that $\tau_2=10^{-5}$, and (b)
$\tau_2=10^{-3}$. The optical activity is the darker curve, and
the wavelength is expressed in units of $\xi_R=25$nm.}
\label{Fig:AOvslambda}
\end{figure}

We now consider the complete wavelength dependance of the optical
activity for the modes $m=\pm 2$. In Figs.~\ref{Fig:AOvslambda}a
and \ref{Fig:AOvslambda}b, the optical activity and the circular
dichroism are shown as a function of the wavelength expressed in
units of $\xi_R$, as calculated with Eqs.~(\ref{OAG}) and
(\ref{Im_Sigma}). In Fig.~\ref{Fig:AOvslambda}a $\tau_2=10^{-3}$
and in Fig.~\ref{Fig:AOvslambda}b $\tau_2=10^{-5}$. Note that in
all these figures, a positive value of $\kappa$ has been chosen
corresponding to a right-handed helix in the cholesteric phase,
but that an opposite optical activity and circular dichroism would
have been found with a left-handed helix. These figures clearly
show a maximum in the magnitude of the optical activity when the
wavelength is equal to the pitch $P$ of the cholesteric phase,
which is about $62.8 \xi_R$ for $\kappa=0.2$ in the case of
Fig.~\ref{Fig:AOvslambda}. This maximum at this wavelength is the
remains in the isotropic phase of the divergence present in the
optical activity of BP I and BP II and in the de Vries formula in
Eq.~(\ref{Vries_f}). As $t$ gets closer to $t_2$, $\Gamma_2(q)$
becomes peaked at $q=\kappa$, and the width of the maximum in the
magnitude of the optical activity gets smaller as can be seen by
comparing Figs.~\ref{Fig:AOvslambda}a and \ref{Fig:AOvslambda}b.

These figures also show the circular dichroism which to our
knowledge has never been discussed theoretically or measured
directly in isotropic chiral media. The absence of experiments is
probably due to the difficulty separating the contribution from
pretransitional fluctuations and the contribution from the
absorption bands. According to Fig.~\ref{Fig:AOvslambda}, circular
dichroism is best observed in the region where $\omega \gg \kappa$
as it gets very small when $\omega \ll \kappa$. The curve shown in
the figure is the contribution from pretransitional fluctuations,
which is valid away from absorption bands. Note that
experimentally this circular dichroism translates in the multiple
light scattering regime into a difference in the scattering mean
free paths for circularly polarized waves. This means that the
transport properties of light in the multiple light scattering
regime should be different for the different states of circular
polarization. Such a difference has indeed been observed in static
and dynamic light scattering measurements with circularly
polarized light in the isotropic phase and in BP III
\cite{collings6}.

\begin{figure}
{\par\centering
\resizebox*{4.6in}{3.2in}{\rotatebox{0}{\includegraphics{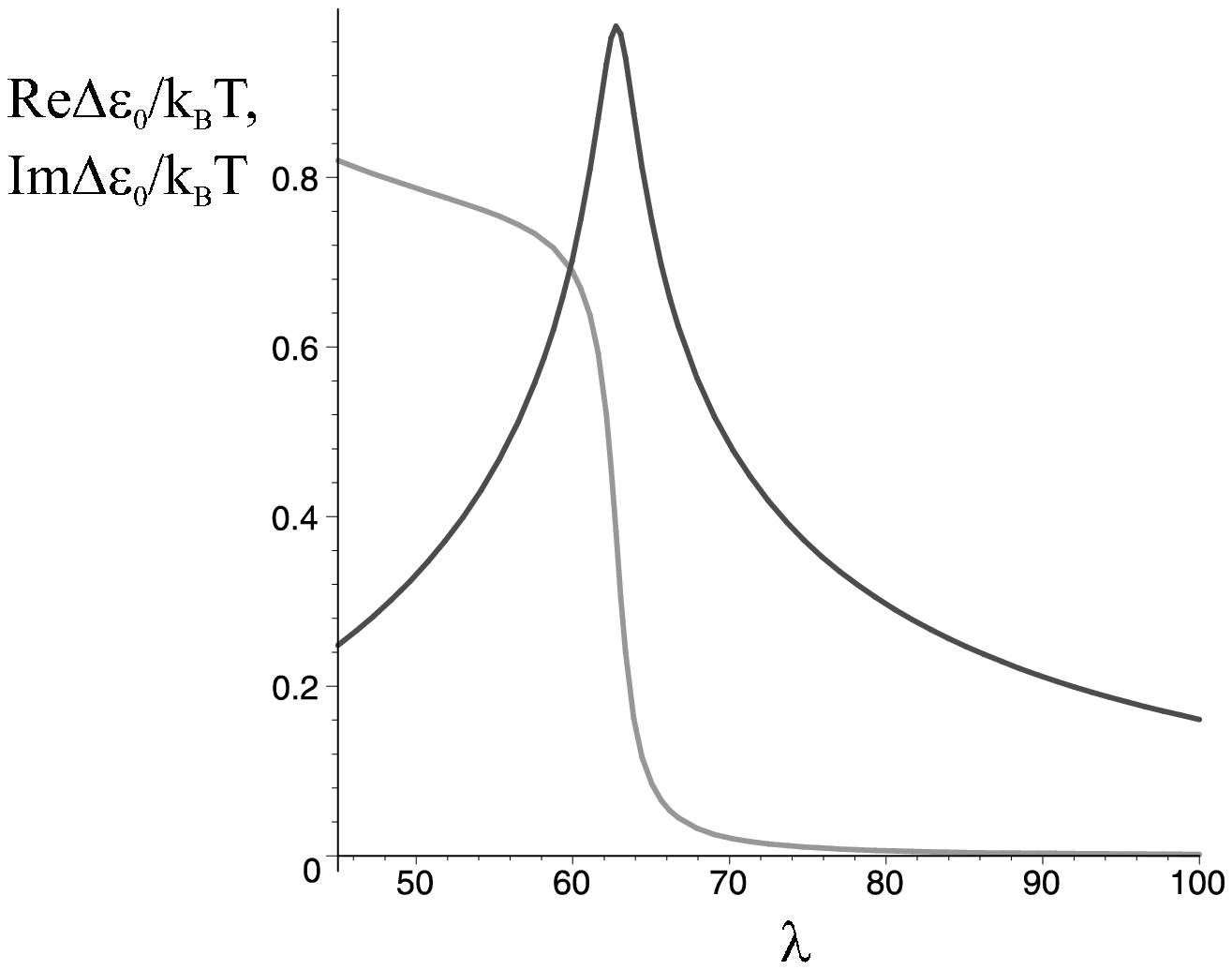}}}
\par} \caption{Real (in dark) and imaginary (in light) part of the $m=\pm2$ contribution
to the average dielectric constant $\Delta \epsilon_0$ as a
function of the wavelength which is expressed in units of
$\xi_R=25$nm. Note that $\kappa=0.2$ and $\tau_2=10^{-5}$.}
\label{Fig:indice}
\end{figure}

Finally we have applied our method to the modification of the
average dielectric constant due to the fluctuations. In section
\ref{sec:theory}, we have defined this quantity as $\Delta
\epsilon_0=-\Sigma_0(\omega,\omega)/\omega^2$, where $\Sigma_0$ is
the isotropic part of $\Sigma$. The effect of the fluctuations on
the symmetric part of the dielectric constant has to our knowledge
never been calculated or measured in the pretransitional region.
We find that all the five modes contribute to this average
dielectric constant. In order to illustrate a case where the
corrections can be large, we focus here on the contribution of the
modes $m=\pm2$, which as noted before is dominant when $t$ is
sufficiently close to $t_2$. Using Eqs.~(\ref{Sigma}) and
(\ref{def_sigma12}), we find
\begin{equation}\label{Indice}
\Sigma_0^{m=\pm2}(\omega,\omega)=\frac{\omega^4}{2 (2\pi)^3} \int
d^3 {\bf q} \frac{ \left( 1+c^2 \right) \Lambda_2(q)}{2q\omega c
-q^2 + i0^+},
\end{equation}
with $\Lambda_2(q)=\Gamma_2(q)+\Gamma_{-2}(q)$. After integrating
over $c$ and $q$, one obtains
\begin{equation}\label{ReIndice}
\re \Delta \epsilon_0^{m=\pm2} = \frac{\omega^3}{64 \pi
\sqrt{\tau_2} \kappa} \left[ g(x_1,t)-g(x_3,t) \right] k_B T,
\end{equation}
with \begin{equation} g(x,t)=-\frac{t+(2\omega x)^2}{\omega^2}
\left[ \left( 1+x^2 \right) \ln \left( \frac{-1+x}{1+x} \right)
+2x \right],
\end{equation}
and
\begin{equation}\label{ImIndice}
\im \Delta \epsilon_0^{m=\pm2} = \frac{\omega^3}{4 \pi} \int_0^1
\left( 1+x^2 \right) x dx \Lambda_2(2\omega x).
\end{equation}
As imposed by the symmetry relation of
Eq.~(\ref{chiral_symmetry}), $\Delta \epsilon_0$ of
Eq.~(\ref{ReIndice}) is an even function of $\kappa$. In
Fig.~\ref{Fig:indice}, the real and imaginary part of the average
dielectric constant $\Delta \epsilon_0$ is shown as a function of
the wavelength, as calculated using Eqs.~(\ref{ReIndice}) and
(\ref{ImIndice}) for the modes $m=\pm2$. Similarly to the case of
the optical activity, there is a maximum in $\Delta \epsilon_0$ as
a function of the wavelength when the wavelength is of the order
of the pitch of the cholesteric.

\section{Theory vs. Experiment} \label{sec: Theory&exp}
\begin{figure}
{\par\centering
\resizebox*{4.6in}{3.2in}{\rotatebox{0}{\includegraphics{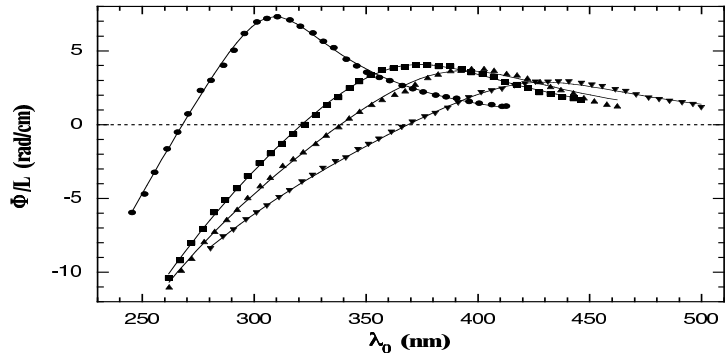}}}
\par}
\caption{Measured ORD spectra (points) in the BP III phase as
function of the wavelength of light in vacuum $\lambda_0$,
together with a fit (solid curve) using Eqs.~(\ref{Mode1}) and
(\ref{OAG}). The ORD spectra are shown for pure cholesteryl
myristate (CM) with circles, for pure cholesteryl nonanoate (CN)
with squares, for mixtures of CN with $5$ and $10$ mol\% of
cholesteryl chloride (CC) with upper and lower triangles,
respectively.} \label{Fig:OA_fit}
\end{figure}

\begin{figure}
{\par\centering
\resizebox*{4.6in}{3.2in}{\rotatebox{0}{\includegraphics{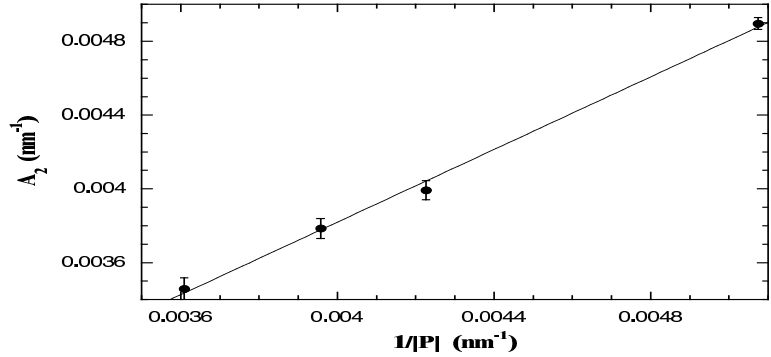}}}
\par}
\caption{Chirality parameter $A_2$ deduced from the fit of
Fig.~\ref{Fig:OA_fit} as a function of the absolute value of the
inverse pitch of the cholesteric phase $1/|P|$.}
\label{Fig:chirality_fit}
\end{figure}

    In this section, we compare the prediction of our model with the
experimental spectra of the optical activity (ORD spectra) in the
BP III phase taken from Ref.~\cite{collings7}. This reference is
the only one known to us that reports measurements of the optical
activity in the isotropic phase and in BP III which are not
limited to the long wavelength regime. The ORD spectra are
measured in the BP III phase for pure cholesteryl myristate (CM),
pure cholesteryl nonanoate (CN) and mixtures of CN with small
quantities of cholesteryl chloride (CC), a compound of opposite
chirality. The sign of the optical rotation agrees with section
\ref{sec:de Vries}: all the samples have a dominant left-handed
character ($\kappa <0$ and $P<0$), and the optical activity is
indeed found to be negative when $\lambda \ll |P|$. The optical
activity presents a broad maximum at $\lambda \simeq -P$, which
grows smaller and broader and shifts to higher wavelength as $|P|$
is increased. The curves are fitted with a theoretical expression,
which is the sum of Eq.~(\ref{Mode1}) (for the modes $m=\pm1$) and
Eq.~(\ref{OAG}) (for the modes $m=\pm2$), and as can be seen in
Fig.~\ref{Fig:OA_fit}, the fits are excellent. The spatial
dispersion of the index, which affects the wavelength of the light
in the medium $\lambda$, is incorporated in this fit using the
measurements on CN reported in Ref~\cite{pelzl}. There are four
free parameters in this fit $A_1$,$A_2$,$A_3$ and $A_4$. These
parameters are $A_1=\pi^2 k_B T \xi_R^3 /8\sqrt{\tau_2}$,
$A_2=-\kappa/4\pi \xi_R=1/|P|$, $A_3=\sqrt{\tau_2}/4\pi \xi_R$,
and $A_4=\kappa k_B T \xi_R^2 \pi / 12\sqrt{\tau_1} r^{3/2}$ so
that $x=(\pm A_2+iA_3)\lambda$ where $\lambda$ is expressed in nm
so that $A_2$ and $A_3$ have units of nm$^{-1}$. The values of
these parameters for the different samples are shown in Table
\ref{tab:coeff}. The parameter $A_2$ obtained from the fits should
be identical to the absolute value of the inverse pitch of the
cholesteric phase. This is indeed the case as shown in
Fig.~\ref{Fig:chirality_fit}, where the points correspond to the
measurements of Fig.~\ref{Fig:OA_fit} and the solid line is a
linear fit which gives a slope of $0.98\pm0.04$ and an intercept
of $(-1.2\pm1.6) \cdot 10^{-4}$.

\begin{table}[h]
  \centering
{\par\centering
\resizebox*{6.5in}{1.2in}{\rotatebox{0}{\includegraphics{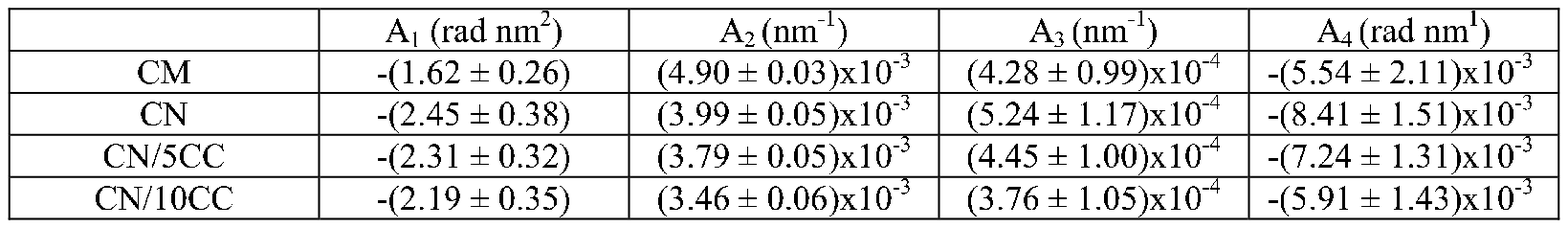}}}
\par}
 \caption{Parameters of the fit of the optical activity spectra for the different
 samples: pure cholesteryl myristate (CM),
pure cholesteryl nonanoate (CN) and mixtures of CN with 5 and 10
mol$\%$ of cholesteryl chloride (CC).}\label{tab:coeff}
\end{table}

\section{Conclusion}
In this article, we have discussed spatial dispersion effects
(optical activity, circular dichroism and average index) for light
propagation in the isotropic-BP III phase, by introducing an
effective index which describes the fluctuations. We have obtained
the wavelength dependance of the spatial dispersion effects
without relying on the long wavelength approximation, on which
previous studies on this problem have been based. This new
approach allows us to discuss the optical properties in the
resonant region when the wavelength of the light is of the order
of the pitch of the cholesteric. In the vicinity of this point,
there is a divergence of the magnitude of the optical activity in
the cholesteric, BP I, and BP II phases and a broad maximum in the
case of the BP III and isotropic phases. These features are
confirmed by measurements of spectra of the optical activity in
the BP III phase for different values of the chirality. We have
also provided predictions for the wavelength dependance of the
circular dichroism and for the symmetric part of the effective
index. We hope that this work will motivate experimentalists to
study the
optical properties of periodic and non-periodic chiral media.\\

We acknowledge many stimulating discussions with B. Pansu, P.
Ziherl and P. Galatola. This work was supported in part by the
MRSEC program under grant NSF DMR00-79909. D. Lacoste received
support by a grant from the French Ministry of Foreign Affairs.

\appendix
\section{Derivation of $\Sigma_1(k,\omega)$}
\label{AppendixA} Using Eqs.~(\ref{Sigma}) and
(\ref{Delta_Sigma}), we find that
\begin{equation}\label{Sigma_n}
\frac{\Sigma_1(k,\omega)}{\omega}=\frac{e_{ijm}k_m}{4ik^2} \int
\dq \left[ B_{ijkl}(-{\bf q}) - B_{ijkl}({\bf q}) \right]
G_{kl}^0({\bf k+q},\omega).
\end{equation}
For the modes $m=\pm2$, $B_{ijkl}({\bf q})=\omega^4
 \Gamma_2(q) T_{ik}^2 (\q) T_{jl}^{-2}
(\q)=\omega^4 \Gamma_2(q) m_i m_k m_j^* m_l^*$ according to
Eqs.~(\ref{tensor_B}) and (\ref{def_T2}). Using the relation
\begin{equation}\label{Relation1}
e_{ijl} k_l m_i m_j^*=-i \k \cdot \q,
\end{equation}
together with Eq.~(\ref{m}), Eq.~(\ref{Sigma_n}) takes a simpler
form
\begin{equation}\label{Sigma_12}
\frac{\Sigma_1^{m=\pm2}(k,\omega)}{\omega}=\frac{\omega^4 \k \cdot
\q}{4k^2} \int \dq \Gamma_2(q) \left( m_k^* m_l + m_k m_l^*
\right) G_{kl}^0({\bf k+q},\omega).
\end{equation}
With the definition Eq.~(\ref{Green_libre}), this expression
reduces to Eq.~(\ref{Vries}).

Similarly, $B_{ijkl}({\bf q})=\omega^4
 \Gamma_1(q) T_{ik}^1 (\q) T_{jl}^{-1}
(\q)=\omega^4 \Gamma_1(q) \left( {\hat q}_i m_k + m_i {\hat q}_k
\right) \left( {\hat q}_j m_l^* + m_j^* {\hat q}_l \right)$ for
the modes $m=\pm1$ according to Eqs.~(\ref{tensor_B}) and
(\ref{def_T1}). Using the relation
\begin{equation}\label{Relation2}
e_{ijl} m_i {\hat q}_j k_l = i \k \cdot \m,
\end{equation}
Eq.~(\ref{Sigma_n}) takes the form
\begin{equation}\label{Sigma_11}
\frac{\Sigma_1^{m=\pm1}(k,\omega)}{\omega}=\frac{-\omega^4}{2k^2}
\int \dq \Gamma_1(q) \left( \k \cdot \m^* \, m_k {\hat q}_l + \k
\cdot \m \, {\hat q}_k m_l^* -\k \cdot \q \, {\hat q}_k {\hat q}_l
\right) G_{kl}^0({\bf k+q},\omega),
\end{equation}
which after simple algebra reduces to Eq.~(\ref{noVries}).

\section{Derivation of Eq.~(\ref{OAG})} \label{AppendixB}
Using the change of variables $x=q/2\omega$, Eq.~(\ref{Vries2})
can be written as
\begin{equation}\label{ReS}
\frac{\Sigma_1^{m=\pm 2}(\omega,\omega)}{\omega}=\mp
\frac{\omega^4}{8 \pi^2} \int_0^{\infty} x dx \Gamma_{\pm 2}(2
\omega x) K(x),
\end{equation}
where $K(x)$ denotes the integral
\begin{equation}\label{K}
K(x)=\int_{-1}^{1} \frac{c^2 \left(1+c^2 \right) dc}{ (x-i0^+
-c)(x-i0^+ +c)},
\end{equation}
which can be decomposed in
\begin{equation}\label{ReK}
\re K(x)=-2x^2-\frac{8}{3}+(x+x^3)\ln \frac{|x+1|}{|x-1|},
\end{equation}
and
\begin{equation}\label{ImK}
\im K(x)= \pi x \left( 1+x^2 \right),
\end{equation}
if $-1 \leq x \leq 1$ and $0$ otherwise. The integration over $q$
in Eq.~(\ref{ReS}) can be carried out in the complex plane with
the function $f$ defined in Eq.~(\ref{function_f}) which is
identical with $\re K$ except for the absence of absolute values
and for this reason $f$ is analytical. $\Gamma_2(q)$ has only two
poles in the upper half plane $\kappa+i \sqrt{\tau_2}$ and
$\kappa-i \sqrt{\tau_2}$, which correspond to $x_1$ and $x_3$ in
Eq.~(\ref{OAG}).

Similarly, the imaginary part of
$\Sigma_1^{m=\pm2}(\omega,\omega)$ can be obtained from
Eqs.~(\ref{ReS}) and (\ref{ImK})
\begin{equation}\label{Im_Sigma}
\im \frac{\Sigma_1^{m=\pm2}(\omega,\omega)}{\omega}=\mp
\frac{\omega^4}{8\pi} \int_0^1 x dx \Gamma_{\pm2}(2\omega x) x
(1+x^2),
\end{equation}
which can also be calculated analytically as a function of
$\tau_2$, $\kappa$ and $\omega$.

%\bibliographystyle{unsrt}
%\bibliography{bp}

\end{document}